\documentclass[onecolumn,prb]{revtex4}
\usepackage[T1]{fontenc}
\usepackage[utf8]{inputenc}
\usepackage{amsmath,amssymb}
\usepackage{color}
\usepackage{graphicx}
\usepackage{amssymb}
\begin{document}
\title{Flat Bands in Slightly Twisted Bilayer Graphene}
\author{E. Su\'{a}rez Morell, J. D. Correa, P. Vargas, M. Pacheco* and Z. Barticevic  }
\affiliation{Departamento de F\'{i}sica, Universidad T\'{e}cnica
Federico Santa Mar\'{i}a, Casilla 110-V, Valpara\'{i}so Chile}
\email{monica.pacheco@usm.cl}

\begin{abstract}
The bands of graphite are extremely sensitive to topological defects which modify the electronic structure. In this paper we found non-dispersive flat bands no farther than  10 meV of the Fermi energy in slightly twisted bilayer graphene as a signature of a transition from a parabolic dispersion of bilayer graphene to the characteristic linear dispersion of graphene. This transition occurs for relative rotation angles of layers around $1.5^o$ and is related to a process of layer decoupling.  
 We have performed ab-initio calculations to develop a tight binding model with an interaction Hamiltonian between layers that includes the $\pi$ orbitals of all atoms and takes into account interactions up to third nearest-neighbors within a layer.
\end{abstract}

\maketitle


Graphene is a sheet of carbon atoms arranged on a honeycomb lattice recently obtained by micro-mechanical cleavage of graphite\cite{key-1}. This two-dimensional lattice  has a singular linear band  dispersion \cite{key-2} which makes the charge carriers behave as massless Dirac fermions with a speed of $\textit{v}_{F}\approx 10^6 m/s$, travelling large distances without interaction. These properties  change drastically in the presence of a second layer. A Bernal or AB stacked bilayer graphene (BLG) is a stack of two carbon sheets in such a way that an atom of one of the layers is in the center of the other layer's hexagon. Like single layer graphene,  BLG  is also a  semimetal but its dispersion relation is quadratic and its charge carriers have a non-zero effective mass.

Few layers of graphene grown epitaxially on different surfaces show a variety of defects including  rotations of the top layer, for instance in graphene layers prepared by chemical vapor deposition (CVD)\cite{key-10}. Moir\'{e} patterns are very often observed in Scanning Tunneling Microscopy (STM) measurements when a relative rotation between top layers is present\cite{key-11,key-12}.

Graphite is also extremely sensitive to topological defects which can modify its electronic structure. Extended defects like lattice dislocations lead to the presence of localized states at the Fermi energy\cite{key-loc1,key-loc2} as is the case of graphene ribbons with zigzag edges\cite{key-loc3}. These localized states can also be found as a result of local defects such as in graphene antidot lattices\cite{key-loc4}. They were predicted in superstructures with honeycomb symmetry by N. Shima et al. \cite{key-loc6}, who suggest the occurrence of ferromagnetism when electron correlation is turned on.

In addition one of the mechanisms proposed to explain high- T$_c$  superconductivity is associated with the presence of extended Van Hove Singularities (VHS)  near the Fermi energy\cite{key-Sup1,key-Sup2, key-Sup3}.  This kind of VHS arising from a nearly dispersionless band  has been observed  in high-T$_c$ cuprates  by angle-resolved photoemission\cite{key-Sup4,key-Sup5}. The superconductivity in graphene when the fermi level is close to a  VHS has been also explored theoretically\cite{key-130}. Recently Guohong Li et al.\cite{key-EvaNat2010} reported the observation of  two symmetric low-energy VHSs in the density of states, measured  by  scanning tunneling spectroscopy, in  twisted graphene layers.  They showed  that the position of these singularities can be tuned  by controlling the relative angle between layers.

In this paper we predict the presence of flat bands, close to the fermi level, in slightly twisted BLG for angles around $1.5^o$.  We show  that for rotation angles below $1^o$,  as expected, the parabolic band dispersion of BLG is obtained, while for angles above $10^o$ up to the symmetric point of $30^o$, the dispersion is linear as in single layer graphene at the K point\cite{key-17,key-18,key-132}. For rotation angles between $1^o$ and $2^o$ our results show a transition to a linear dispersion, otherwise  between $2^o$ and $10^o$, a linear dispersion prevails with a renormalized velocity of Dirac fermions\cite{key-13,key-132}.

A rapid flattening of the bands is observed when we start rotating a stacked AB Bilayer,  flat bands very close to the Fermi level are revealed for angles around $1.5^o$.  This  effect has been reported in Ref.\cite{key-EvaNat2010} were they observe two pronounced peaks in the  DOS measured by STM, and by changing the rotation angle between the layers they achieve a constant shift of the VHS's toward Fermi level. In this work, based on a band structure analysis, we show  the existence of a critical angle for which the energy separation between the VHS's reaches a minimum.

We have developed a tight-binding (TB) model to tackle commensurable unit cells with a large amount of atoms. The model  considers the interlayer interaction of $\pi$ orbitals  of all atoms between layers and up to third nearest-neighbors interaction within a layer. We fit  our TB model to reproduce the band structure obtained from Density Functional Theory  (DFT) calculations of a stacked AB bilayer.

For building a commensurate unit cell we have followed a procedure proposed  by Kolmogorov and Crespi\cite{key-14}. We started our rotations  from a stacked AB bilayer. In this stacking there are two non-equivalent sites,  A ($\alpha$) site, where an atom lies directly above another atom and,  B ($\beta$) site where an atom position is just in the center of another layer hexagon. We have chosen a B site as our rotation center, however the results are essentially the same if we choose an A site, the difference is that with the latter we can obtain a stacked AA superstructure (all atoms in one layer are above/below another).

We do a clockwise commensurable rotation from a vector $\vec{r}=m \vec{a}_{1} + n \vec{a}_{2} $ to  $\vec{t}_{1}=n \vec{a}_{1} + m \vec{a}_{2}$, where $\vec{a}_{1}=(-1/2,\sqrt{3}/2)a_{0}$ and  $\vec{a}_{2}=(1/2,\sqrt{3}/2)a_{0}$ are bilayer lattice vectors; \textit{m},\textit{n} are integers and $a_{0}=2.46 \mathring{A} $ is the lattice constant. The unit cell vectors are : $ \vec{t}_{1}=n \vec{a}_{1} + m \vec{a}_{2} $ and $\vec{t}_{2}= -m \vec{a}_{1} + (n+m) \vec{a}_{2}$ and it contains $N=4(n^{2}+ mn+m^{2})$ atoms.

\begin{figure}
\begin{centering}
\includegraphics[scale=0.5,angle=0]{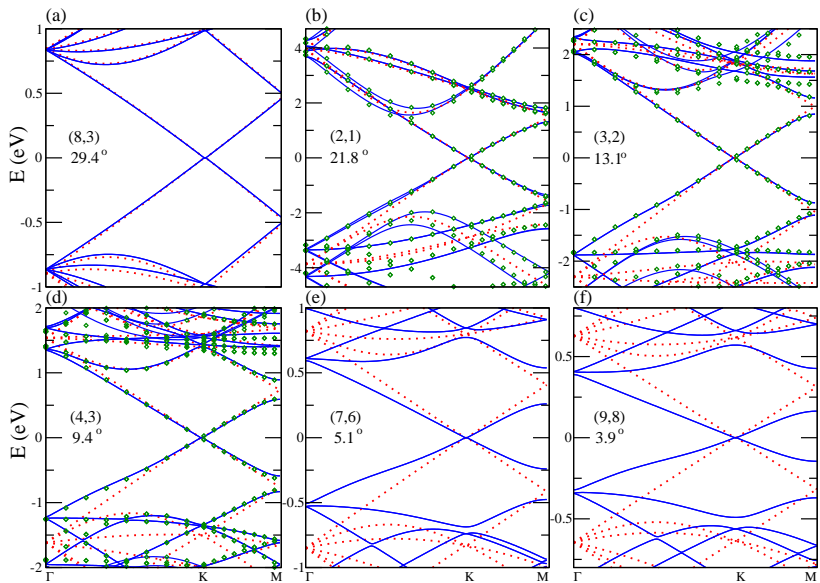}
\par\end{centering}
\caption{(Color Online) Energy band dispersion for 6 different structures. Twisted BLG
(straight blue line),  single layer graphene (red dotted) and DFT results (green diamonds).
At the Fermi level, around the $K$-point a linear behavior of the bands is observed for all structures.  In
\textbf{e} and \textbf{f} the Fermi velocity is renormalized.}\label{fig:1}
\end{figure}

We have adopted the tight binding model proposed by Reich et al.\cite{Reich2002} to study the electronic structure of graphene, which includes third-nearest neighbor approximation for the non-orthogonal  $\pi$ orbitals model. Such
parameterization  properly describes the $\pi$ and $\pi^{*}$ graphene bands and it is able to reproduce first principle results over the entire Brillouin zone.  The Hamiltonian for twisted stacks is
given by $H=H_{1}+H_{2}+H_{int}$, where $H_{1}$ and $H_{2}$ are the Hamiltonians for
each layer following the Reich  model and
$H_{int}=\gamma_{t}e^{-(\mathbf{r}-\mathbf{d})/\beta}. \label{eq:hamil-int}$ is the  interaction Hamiltonian between the  two layers. Here  $\mathbf{d}$ is the interlayer distance and $\gamma_{t}$ and $\beta$ are parameters.

By choosing this interaction Hamiltonian we are taking into account the complexity of the unit cell where the distances between atoms of different layers are all different. There are nine parameters  in this \textit{TB} model and they were obtained fitting the band structure of a stacked AB bilayer to reproduce DFT results. 
 In our calculations the interaction between layers is not restricted to the nearest neighbors,  we have included interaction between all atoms of different layers.  If only the nearest neighbor is included in the interlayer interaction, the AA site gets a higher significance (a unique site in each cell where one atom lies exactly above another), overestimating the renormalization of fermi velocity\cite{key-13}. 



\textit{Ab-initio} calculations of the electronic structure of BLG and
some  twisted configurations (namely (2,1);(3,2);(4,3),(5,4)) were done using the SIESTA\cite{key-221}
code with atomic basis DZP (double-$\zeta$ with polarization), using the \textit{ LDA}\cite{key-222} approximation for the exchange correlation potentials with an energy cut of $E_{cut}=270R_{y}$.
The Monkhorst Pack grid $10\times10\times1$ was used as a k-points grid. All structures were fully  relaxed using  the Conjugate  Gradient
algorithm \textit{CG} with force tolerance of $0.02eV/\textrm{\AA}$.
The interlayer distance used was  $c=3.35 \mathring{A} $.
\begin{figure}
\begin{centering}
\includegraphics[scale=0.5,angle=0]{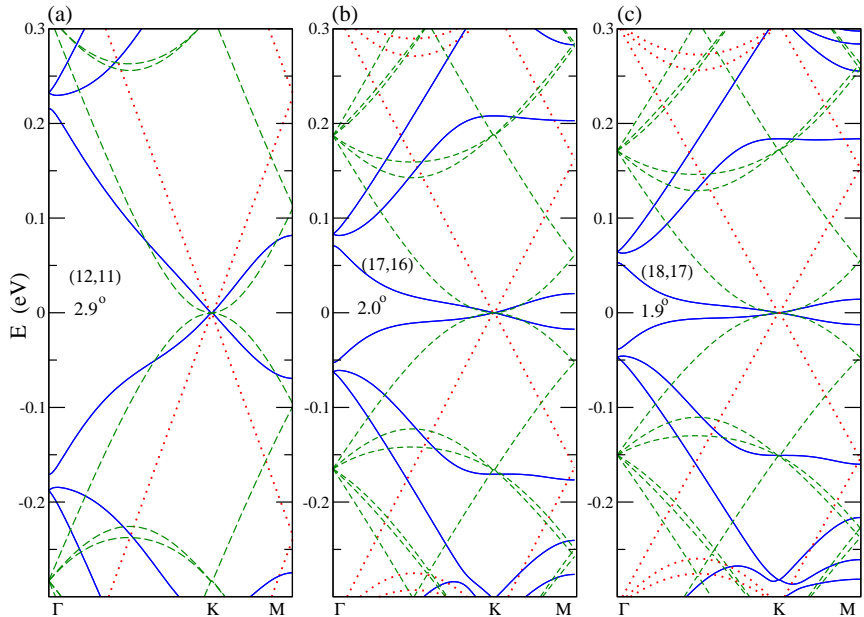}
\par\end{centering}
\caption{(Color online) A flattening of the bands is observed with a depletion in $\textit{v}_{F}$ for the slightly twisted Bilayer (blue line). The AB stacked BLG (green dashed line)  and single layer graphene (red dotted line) are  included for  comparison.}\label{fig:2}
\end{figure}
We have compared the band structures obtained by SIESTA calculations  with those obtained with our TB model for the four rotated stacks mentioned, we found no difference in the low energy range. In Fig.\ref{fig:1} (b),(c) and (d) it is shown the corresponding band structures for different cells calculated with DFT and with the TB model.


In figure \ref{fig:1} we show the behavior of twisted bilayer graphene and single layer graphene for angles between  $3.9^o$ to nearly $30^o$, in six different configurations. The K and M points have different values for each cell, so for comparison we included the one layer and/or bilayer graphene energy bands in every plot, which are calculated with the same unit cell vectors of the twisted bilayer graphene.

 At the Fermi level and around the K point, a linear band  dispersion of the twisted cell can be seen in all graphics. In Fig.\ref{fig:1}(a), (b), (c) and \ref{fig:1}(d), the slope of the dispersion curves is identical to those of a single layer  graphene and in Fig.\ref{fig:1}(e) and \ref{fig:1}(f)  a decreasing of the band slope is evident around the K point in the rotated BLG.
\begin{figure}
\begin{centering}
\includegraphics[scale=0.5,angle=0]{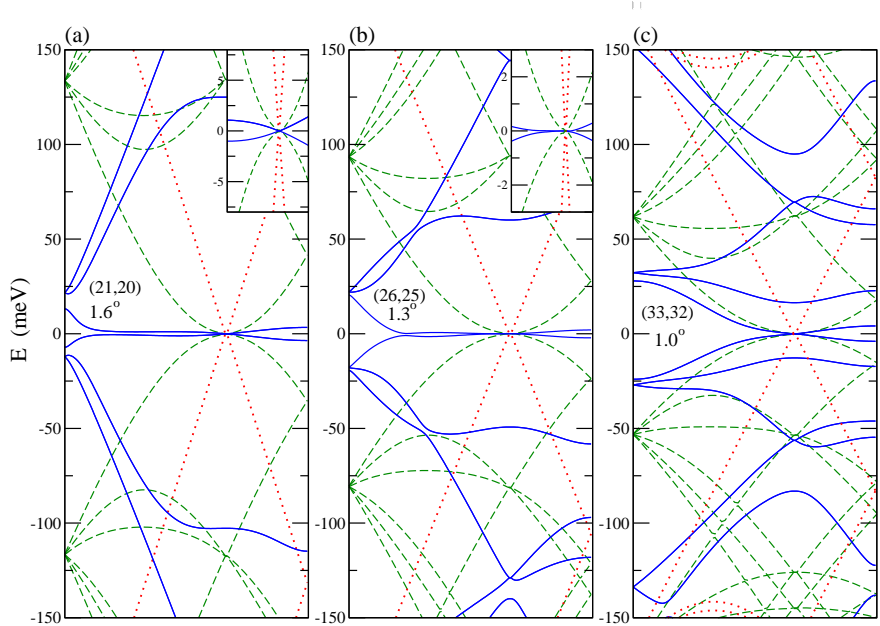}
\par\end{centering}
\caption{ (Color online) Localized states in the band structure of twisted BLG for angles around $1.5^o$. Flat bands are not present for $1^o$, instead a parabolic dispersion appears. Insets in \textbf{a}  and \textbf{b} are obtained near the K point. They show a change in the  convexity of the band from $\Gamma$ to K point  and, in \textbf{b}, a near zero Fermi velocity is revealed.}\label{fig:3}
\end{figure}

For rotation angles above $10^o$ the weak coupling between layers  is responsible for this behavior. The Fermi velocity $ \textit{v}_{F} $ at the \textbf{K} point of the rotated bilayer is the same as in single layer graphene in agreement with \cite{key-17,key-18,key-132}. An increase in the  interlayer distance, obtained by DFT relaxation \cite{key-16}, only enhances this effect.
It is important to notice that in  the review published by Pong et al.\cite{key-19} most of the Moir\'{e} patterns have been observed for angles below $10^o$. These structures do not produce Moir\'{e} Patterns strong enough to be detected by STM\cite{key-20}.


In figure \ref{fig:2} the differences in the slope of the twisted structure compared with that of the single layer graphene are more evident. The stacked AB BLG bands are included to enhance visual comparison. It is also clear that bands become  less dispersive for low angles and they touch the $\Gamma$ point closer to the Fermi level. Our results are in agreement with those  obtained by\cite{key-17,key-13} using different approaches.


\begin{figure}
\begin{centering}
\includegraphics[scale=0.5,angle=0]{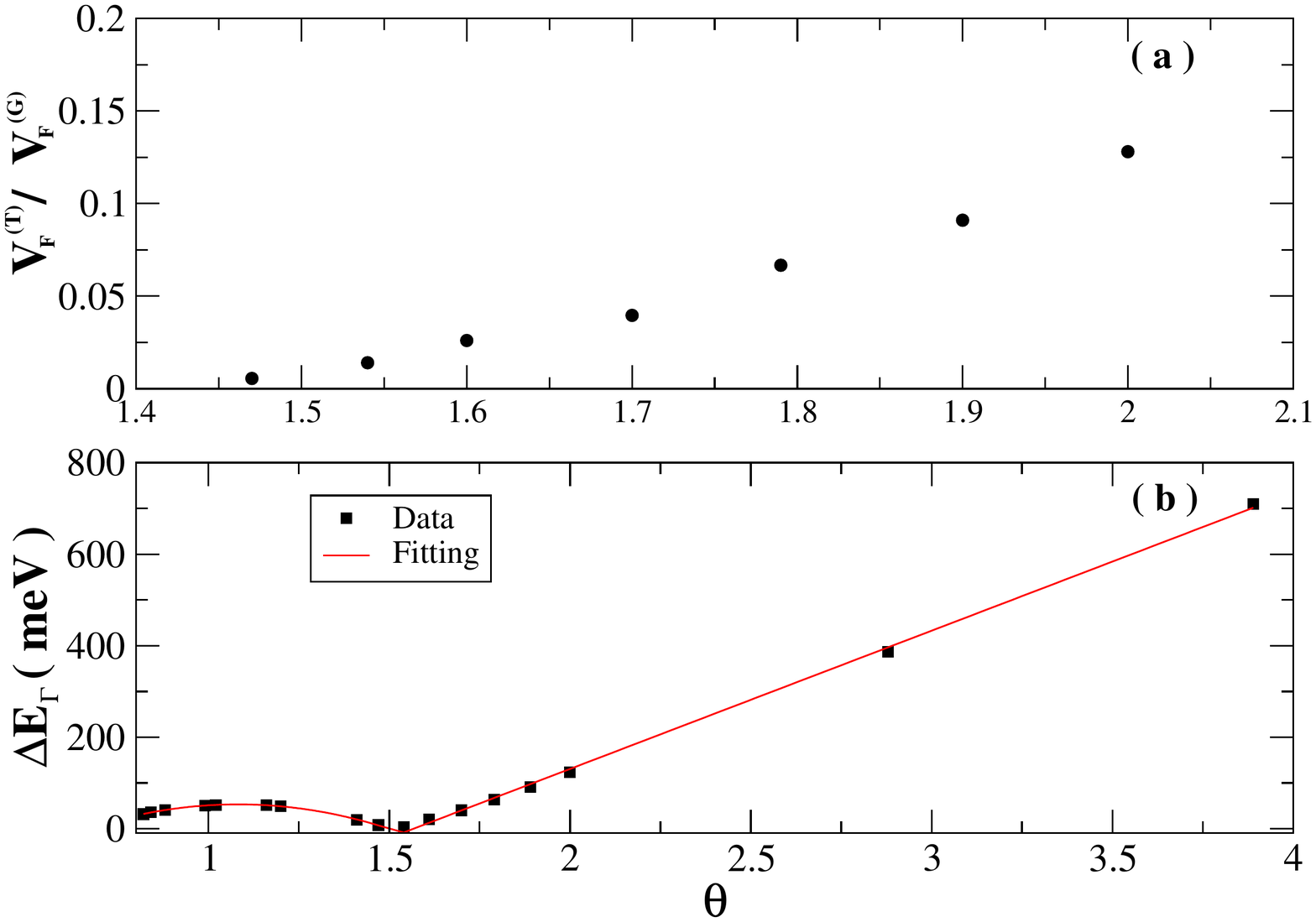}
\par\end{centering}
\caption{(Color online) \textbf{a.} Dependence of Fermi velocity in twisted BLG $\textit{v}_{F}^{(T)}$ and in  single layer graphene $\textit{v}_{F}^{(G)}$ with the rotating angle. \textbf{b.} $\Delta$E$ $ at $\Gamma$ point versus the rotation angle.
}\label{fig:4}
\end{figure}

For relative rotation angles of $1.6^o$ and $1.3^o$ (Fig.\ref{fig:3}(a),(b)) the bands are completely flat. The inset in \ref{fig:3}(b) reveals a near zero Fermi velocity near K point\cite{key-132}. By comparing the insets of Fig.\ref{fig:3} (a) and Fig.\ref{fig:3}(b) a change in the band convexity and the presence of asymmetric flat bands very close to the Fermi level can be observed. Otherwise, for an angle of $1^o$ (Fig.\ref{fig:3}(c) the behavior is parabolic as expected.

Since BLG has a parabolic dispersion, we would not expect a drastic change of the band structure if we twist it by a very small angle. By increasing the angle, the layers start a decoupling process and this influences  the band dispersion. Around $1.5^o$ a transition characterized by flat bands occurs. The energy of these bands changes with the angle and so does the relative extension of the flat area. These flat bands are only present for a small range of angles and the separation between them, near the Fermi energy, reaches a minimum at $\theta_{c}=1.5^o$. The relative extension of the flat bands is maximized at the same angle. This latter behavior can be associated with localized states near fermi level.

The  behavior shown by the flat bands mentioned in the preceding paragraph is also observed for the energy gap $\Delta E$ (energy difference between the minimum of the conduction band and the maximum of the valence band) at the $\Gamma$ point. In Fig.\ref{fig:4}(b) we have plotted  this energy gap as a function of the rotation angle. It is clearly seen in the graphic the existence of a critical angle that separates two different behavior of this energy gap. We have fitted the results by separating the data above  and below the critical angle $\theta_{c}=1.5^o$ and we can observe a linear behavior of $\Delta E$  as a function of $\theta$ for angles greater than $\theta_c$ and a quadratic behavior, for smaller angles. This transition point indicates two qualitatively different electronic behavior, a massive particle for small rotation angles between the graphene layers and massless Dirac fermions when the angle exceeds the critical value. A slight shift in the value of the critical angle is quite possible provided the descriptive character of the TB results.

The fact that the behavior of $\Delta$E$ $  at $\Gamma$  point is an indication of the presence of flat bands in Twisted BLG is very convenient because we only need to calculate one point ($\Gamma$) of the band structure to plot this graph, otherwise it would be very time consuming  since unit cells for angles below $1^o$ contain more than 15000 atoms.

 In Fig.\ref{fig:4}(a) we also plotted the quotient between the Fermi velocity in twisted BLG  and the corresponding Fermi velocity in  single layer graphene, $\textit{v}_{F}^{(T)}/\textit{v}_{F}^{(G)}$, as a function of the twisting angle. By fitting the data we have found that this velocity is depleted for an angle of $\simeq 1.5^o$. For smaller  angles a parabolic dispersion it is retained and thus regarding the presence of Dirac electrons it is not adequate. They are only present for angles above this critical angle.

We believe that the arising of these flat bands in the twisted structure is related to a lost of a degree of freedom in the system, when layers decouple the electrons are confined to each layer, this transition would cause the appearance of extended VHS.


In conclusion we have studied the band structure of a slightly twisted bilayer graphene by employing ab-initio calculations to develop a parameterized tight binding(TB) model.  This model has allowed us to deal with large unit cells and also to compare our results with those previously reported. Our  Hamiltonian includes all interactions between atoms on different layers and takes into account interactions up to third nearest-neighbors within a layer.  We have found the presence of flat bands at the K point around a critical rotation angle of $1.5^o$. This is
the signature of a transition from a parabolic to linear dispersion in the twisted structure.  A degree of freedom is lost in this transition, layers decouple and electrons are confined to each layer for angles above the critical one, and the system behaves like separated graphene layers. For this critical angle the Fermi velocity in a twisted BLG is completely depleted. Therefore,  the presence of Dirac fermions in twisted bilayer graphene is valid only for  angles larger than the critical one. Why this transition occurs precisely at this angle of $1.5^o$ is probably dependent upon our parametrization model, but we have found that a critical angle certainly exists.

\begin{acknowledgements}
E.S.M acknowledges CONICYT(Chile) and UTFSM for the internal grant PIC-DGIP. P.V. acknowledges FONDECYT grant 1100508. M.P. acknowledges  the financial support of USM 110971 internal grant and FONDECYT grant 1100672.
\end{acknowledgements} 


\end{document}